\documentclass[12pt,preprint]{aastex}

\usepackage{epsfig}
\usepackage{natbib}                

        \shorttitle{HD 42401}
        \shortauthors{Williams}

\begin{document}


\title{System Parameters for the Eclipsing B-star Binary HD 42401}

\author{S. J. Williams\footnote{Visiting Astronomer, Kitt Peak National 
Observatory, National Optical Astronomy Observatory, which is operated by the 
Association of Universities for Research in Astronomy (AURA) under cooperative 
agreement with the National Science Foundation.}}
\affil{Center for High Angular Resolution Astronomy and \\
 Department of Physics and Astronomy,\\
 Georgia State University, P. O. Box 4106, Atlanta, GA 30302-4106; \\
 swilliams@chara.gsu.edu}

\begin{abstract}

I present results from an optical spectroscopic investigation of the
binary system HD~42401 (V1388 Ori; B2.5 IV-V + B3 V). A combined analysis of 
$V-$band photometry and radial velocities indicates that the system has an 
orbital period of 2.18706 $\pm$ 0.00005 days and an inclination of 75.5 $\pm$
0.2 degrees. This solution yields masses and radii of $M_{\rm 1}$ = 7.42 $\pm$
0.08 $M_{\odot}$ and $R_{\rm 1}$ = 5.60 $\pm$ 0.04 $R_{\odot}$ for the primary
and $M_{\rm 2}$ = 5.16 $\pm$ 0.03 $M_{\odot}$ and $R_{\rm 2}$ = 3.76 $\pm$ 0.03
$R_{\odot}$ for the secondary. Based on the position of the two stars plotted
on a theoretical H-R diagram, I find that the age of the system is 
$\gtrsim$~25 
Myr and that both stars appear overluminous for their masses compared to
single star evolutionary tracks. A fit of the spectral energy distribution
based on photometry from the literature yields a distance to HD~42401 of 832
$\pm$ 89 parsecs.

\end{abstract}
\keywords{binaries: spectroscopic -- stars: early-type -- stars: fundamental
parameters -- stars: individual (HD 42401)}


\section{Introduction}

Two fundamental parameters in stellar astrophysics are the masses and radii
of stars. Especially lacking are accurate ($\le2\%$) determinations of 
these quantities for O- and B-type systems. Data of this kind are important
for tests of stellar models, with far-reaching implications such as the 
modeling of stellar components of galaxies. Here I describe measurements of
these parameters for the B-star binary HD~42401. This effort is part of an 
ongoing program to determine the masses and radii of O- and B-type stars. 
Previously, I obtained parameters for the LMC O-star binary [L72]~LH~54-425 
\citep{wil08} and will continue with more analyses of eclipsing spectroscopic 
binary systems containing O- and B-type stars.

The star HD~42401 was first classified as B2 V by \citet{wal71}. Since 
then, it has been used as a spectral standard for the B2 V type stars in a 
number of publications \citep{wal90,de00,bag01}. The eclipsing nature of its 
light curve was first noticed with the \textit{Hipparcos} satellite 
(HIP 29321), where it was 
classified as an Algol-type eclipsing binary and given the moniker V1388 Ori 
\citep{kaz99}. Three radial velocity measurements of HD~42401 by \citet{feh97}
show a range from --103 to 36 km s$^{-1}$, but no further investigations were
made into this probable velocity variable. 

Presented here are the analyses of radial velocities from optical spectra 
obtained in 2008 January~(\S 2). Section 3 contains a description of the
measurement of radial velocities from the data. In \S 4 I describe tomographic
reconstructions of the individual component spectra. Covered in \S 5 is the
combined light curve and radial velocity curve solution. In \S 6 I 
finish with a discussion of the distance to the system, its fundamental
parameters, and other consequences of the analysis. 


\section{Observations}

I obtained 29 spectra of HD~42401 with the Kitt Peak National Observatory
(KPNO) 0.9-m coud\'e feed telescope during an observing run from 2008 
Jan 09 to 2008 Jan 18. These spectra were obtained with the long collimator and
grating A (632 grooves mm$^{-1}$ with a blaze wavelength of 6000 \AA~in 
second order with order sorting filter 4-96). The first two nights of
observations were made with the T1KB detector (1024$\times$1024 pixel array 
with 24 square $\mu$m pixels) and resulted in a resolving
power of $R = \lambda/\delta\lambda = $ 13,500 with wavelength coverage
from 4315--4490 \AA. The rest of the observing run made use of the F3KB 
detector (3072$\times$1024 pixel array with 15 square $\mu$m pixels) 
that resulted in a lower resolving power
of $R = \lambda/\delta\lambda = $ 11,500 with wavelength coverage from
4250--4570~\AA. The wavelength coverage was chosen to include the 
He~\textsc{i}~$\lambda$4471 and Mg~\textsc{ii}~$\lambda$4481 lines that are
good temperature indicators for B-type stars plus the H$\gamma$ line that is
sensitive to gravity (linear stark effect). 
Exposure times were 1200 s giving a S/N $\simeq$ 
100 pixel$^{-1}$. Numerous comparison spectra were obtained for wavelength
calibration, and many bias and flat field spectra were also obtained each 
night. The spectra were extracted and calibrated using standard routines
in IRAF\footnote{IRAF is distributed by the National Optical Astronomy
Observatory, which is operated by the Association of Universities for Research
in Astronomy, Inc., under cooperative agreement with the National Science
Foundation.}, and then each continuum rectified spectrum was transformed to a
common heliocentric wavelength grid in log $\lambda$ increments. 


\section{Radial Velocities}\label{sectrv}

The narrow range of wavelength coverage in our spectra limits the lines that 
may be used for radial velocity analysis. I measured radial velocities from
three lines, He~\textsc{i}~$\lambda$4387, He~\textsc{i}~$\lambda$4471, and 
Mg~\textsc{ii}~$\lambda$4481 via a template-fitting scheme \citep{gie02} that 
measures velocities by using model templates weighted 
by a flux ratio to match both the shifts and line depths in the observed 
spectra. There is no evidence of emission or intrinsic line asymmetries in 
these lines or H$\gamma$~$\lambda$4340. 

The BSTAR2006 grid of stellar models from \citet{lan07} was used to derive 
template spectra. These models are based on the line blanketed, non-LTE, 
plane-parallel, hydrostatic atmosphere code TLUSTY and the radiative transfer
code SYNSPEC \citep{hub88,hub95,hub98}. In finding templates, initial values
were used for temperatures, gravities, projected rotational velocities, and
flux contributions from each star. These parameters for model templates were
then checked by comparing the three lines used in radial velocity 
measurements against the tomographically reconstructed spectra of each star
(\S\ref{tomog}). The parameters were changed and new templates made 
after initial fits to the light curve and radial velocity curves (\S 5) 
indicated slightly different values were more appropriate. The velocity 
analysis was then performed again until the best fit was obtained. In this way,
I was able to obtain an estimate of the monochromatic flux ratio
in the blue spectra of $F_{\rm 2}/F_{\rm 1} = 0.25 \pm 0.05$ based upon the
relative line depths of the spectral components.

\begin{figure}
\begin{center}
{\includegraphics[angle=90,height=12cm]{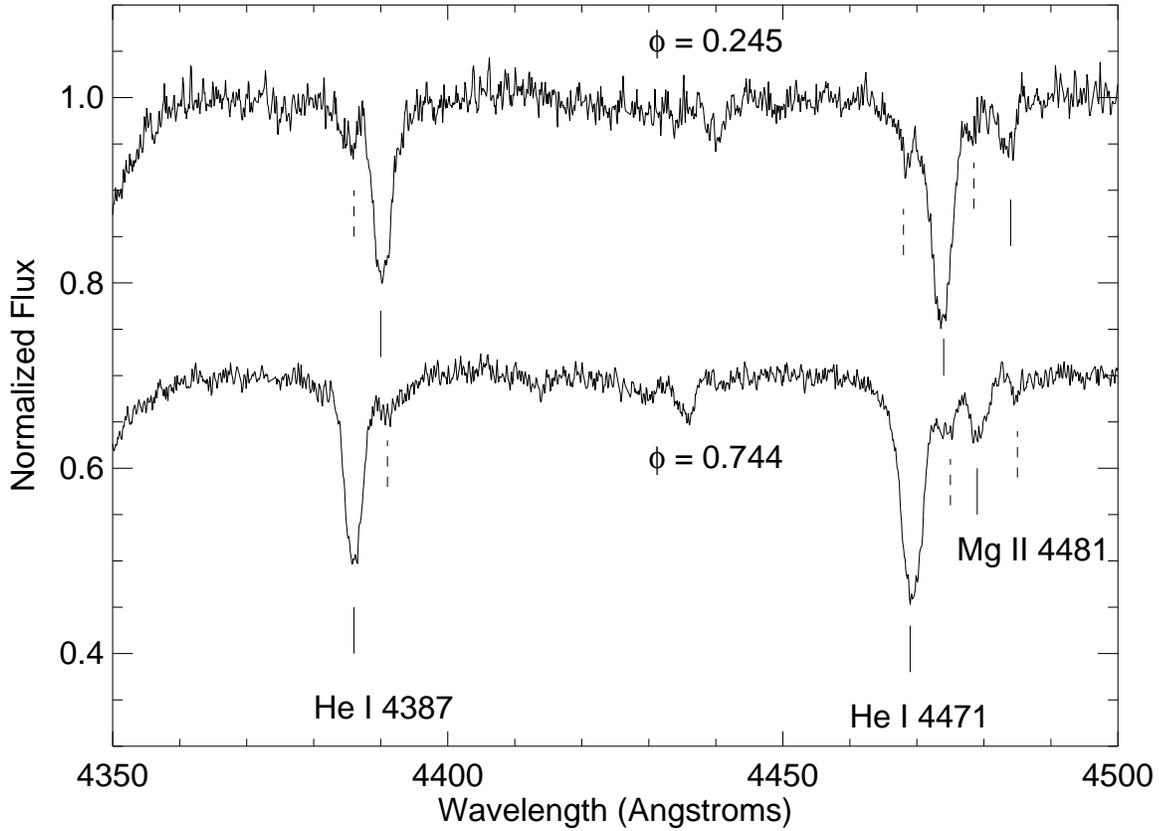}}
\end{center}
\caption{Two spectra of HD~42401 obtained near quadrature phases (offset for 
  clarity). Solid lines indicate the positions of lines of the 
  primary component and dashed lines indicate those for the secondary 
  component. The lines marked are those used in the velocity analysis.}
\label{fig1}
\end{figure}

The template fitting scheme also needs preliminary estimates for the velocities
of each component. To obtain these, spectra which clearly showed two sets of
spectral lines were analyzed. Midpoints of spectral features were used to 
make crude estimates of radial velocities for each star. Preliminary orbital
parameters were obtained by using the nonlinear, least-squares fitting
program of \citet{mor74}. The preliminary velocities from this initial 
orbital solution were starting points for performing a nonlinear, least-squares
fit of the composite profiles with the template spectra and calculating the 
shifts for each star. The fitting scheme searches a region of $\pm$10 
\AA~around the rest wavelength of each line, so preliminary velocities are mere
starting points and need not be highly accurate. 
The three values for radial velocity from each spectrum
were averaged, and the standard deviation of the mean value was calculated. 
Each of these values are listed in Table \ref{rvs} for the primary and 
secondary stars. Also listed are the orbital phase for each observation and the
observed minus calculated values for each data point. Phase zero is defined
as the time of inferior conjunction of the primary star, $T_{\rm IC,1}$ 
(time of secondary minimum in the light curve). Figure \ref{fig1} shows two 
sample spectra at times of near quadrature according to this ephemeris. 

During eclipse phases, velocity measurements may be affected by the 
Rossiter-McLaughlin effect in which the center of light of the eclipsed star
will appear redshifted on ingress and blue shifted on egress because of the
rotational Doppler shifts of the visible portions. 
This effect will result in narrower line profiles as 
the red and blue shifted regions of the eclipsed star are blocked, and it is
more prominent in broad-lined, rapidly rotating systems. HD~42401 is a short 
period binary having relatively narrow lines, therefore the effect should be 
small. However, since the template method does not account for eclipses, 
deviations from orbital motion near eclipse phases are expected.


\section{Tomographic Reconstruction}\label{tomog}

\begin{figure}
\begin{center}
{\includegraphics[angle=90,height=12cm]{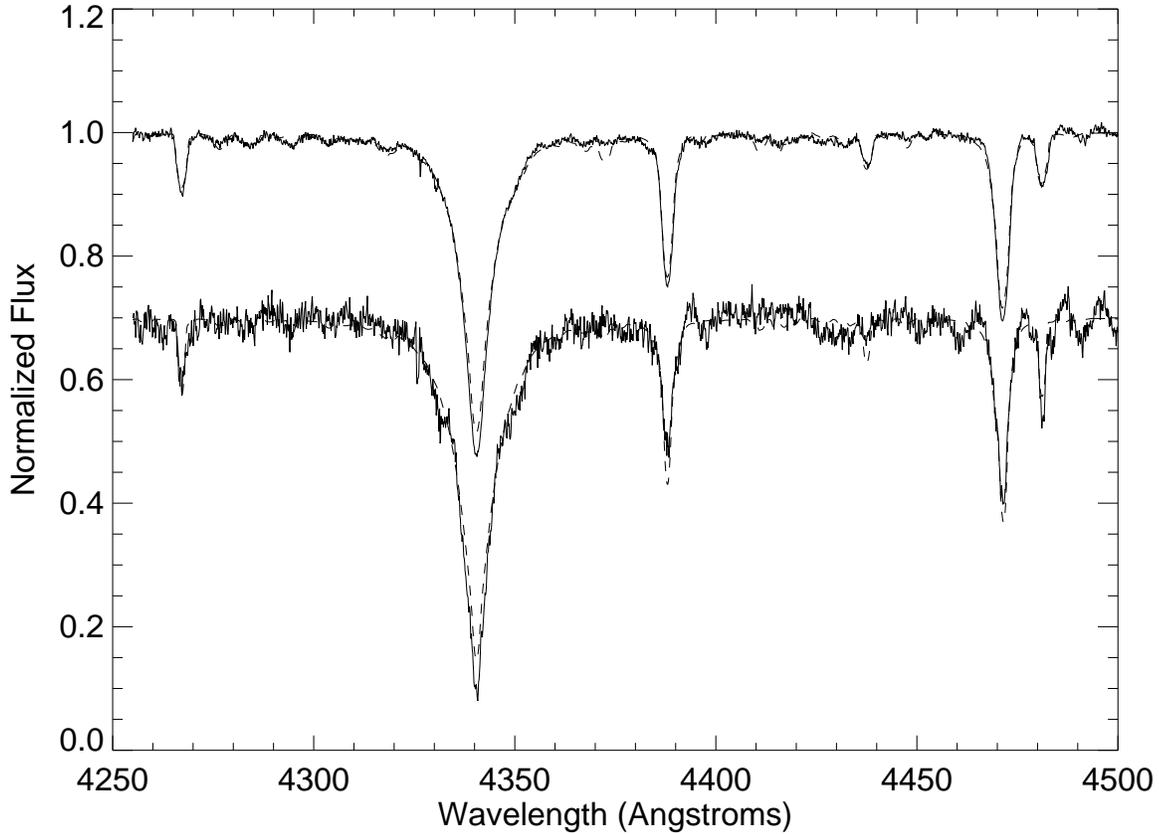}}
\end{center}
\caption{Tomographic reconstructions of the components of HD~42401 based on
  24 spectra obtained with the F3KB instrument during January 2008 at KPNO. 
  The top solid line represents the reconstructed spectrum of the primary, 
  the bottom solid line is the reconstructed secondary spectrum offset by 0.3
  for clarity. Over-plotted for both are the model spectra for each (dashed 
  lines). The stellar parameters for the model spectra are given in 
  Table \ref{tomo}.}
\label{fig2}
\end{figure}

The Doppler tomography algorithm of \citet{bag94} was used to separate the
primary and secondary spectra of HD~42401 for the F3KB spectra. This iterative 
method uses the 24 observed composite spectra, their velocity shifts,
and an assumed monochromatic flux ratio to derive individual
component spectra. The best flux ratio was the one that best matched line 
depths in the reconstructions with those in the model spectra. 
Figure~\ref{fig2} shows the reconstructed spectra for the primary and 
secondary with the best fit models over-plotted. The relative depths of
He~\textsc{i}~$\lambda$4471 and Mg~\textsc{ii}~$\lambda$4481 are good 
temperature indicators throughout the B-star sequence. Specifically, the 
He~\textsc{i}~$\lambda$4471 line gets weaker while the 
Mg~\textsc{ii}~$\lambda$4481 line gets stronger as temperature
decreases, as is seen in the spectrum of the secondary compared to that of
the primary. The set of spectra from the F3KB instrument has only one spectrum
that is close to an eclipse. Because there are many more spectra outside 
eclipse used in the tomographic reconstruction, the final reconstructed spectra
are insensitive to details of the Rossiter-McLaughlin effect. Indeed, 
tomographic reconstructions with the one eclipse spectrum omitted are 
negligibly different from those presented in Fig. \ref{fig2}.

The final reconstructed spectra were fit with TLUSTY/SYNSPEC model synthesis
spectra (see \S \ref{sectrv}). Fits to all three lines made in the velocity
analysis were used to estimate the projected rotational velocity $V \sin i$,
and the temperature and surface gravity of each star were estimated by 
comparing the reconstructed and model profiles for a grid of test values 
\citep[details are given in][]{wil08}. 
These parameter values are listed in Table \ref{tomo}.

The narrow wavelength range of the spectra is insufficient to attempt classical
spectral typing. However, I can arrive at an estimate of the spectral types 
for each 
star in the HD~42401 system by comparing the derived effective temperatures
with a spectral type versus effective temperature relation. 
According to Table 2 of \citet{boh81} the effective temperature and gravity 
(\S 5) of the primary of HD~42401 is most consistent with a B2.5 IV-V star 
while the secondary matches most closely with a B3 V star, and
these classifications are listed in Table \ref{tomo}.


\section{Combined Radial Velocity and Light Curve Solution}

The All Sky Automated Survey \citep[ASAS-3;][]{poj02} database contains light 
curves for $\sim$39,000 previously unknown variable stars. I extracted the
$V$-band light curve for HD~42401 (ASAS 061059+1159.7) from this catalog, 
removing points deemed lower quality by the data reduction pipeline used by 
the ASAS. 

\begin{figure}
\begin{center}
{\includegraphics[angle=90,height=12cm]{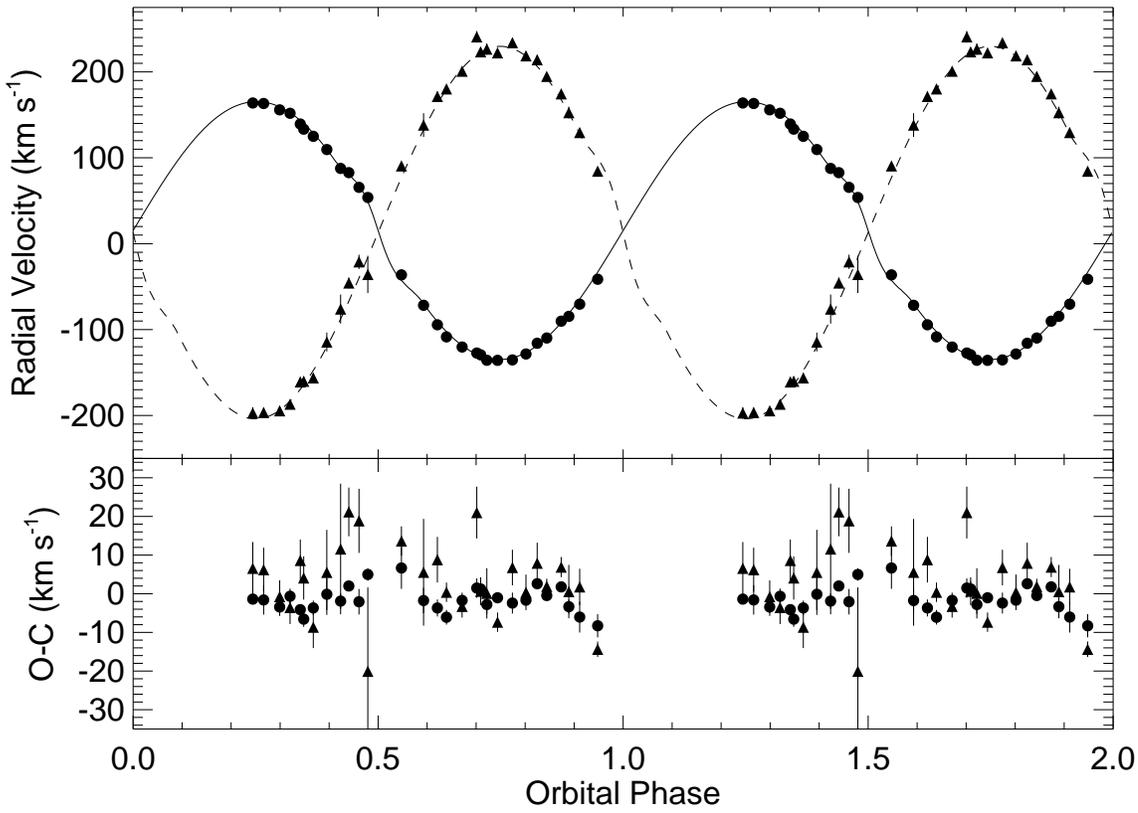}}
\end{center}
\caption{Radial velocity curves for HD~42401. Primary radial velocities are
  shown by filled dots and secondary radial velocities are shown by filled
  triangles with associated uncertainties shown as line segments for both. 
  The solid line is the best fit solution for the primary and the dashed line
  is the same for the secondary. The lower panel shows the observed minus 
  calculated values for each measurement with uncertainties.}
\label{fig3}
\end{figure}

I used the Eclipsing Light Curve (ELC) code \citep{oro00} to find orbital 
and astrophysical parameters for the HD~42401 system. ELC fits the radial 
velocity and light curves simultaneously, giving a joint orbital ephemeris 
based on both sets of data. The best fit for the radial velocity curve is shown
in Figure \ref{fig3}(upper panel) along with the observed minus calculated 
residuals (lower panel). ELC treats calculation of velocities during eclipse 
via a flux-weighted velocity centroid as described in \citet{wil76}. 
The Rossiter-McLaughlin effect is clearly seen in the 
best fit as a slight redshift going into eclipse and a slight blueshift 
leaving eclipse and the model velocity curves match the data well. 
The best fit for the $V$-band 
light curve is shown in Figure \ref{fig4}. Figures \ref{fig5} and \ref{fig6} 
show the details of the fit for the light curve in the region of the primary 
and secondary eclipses, respectively.

The genetic optimizer mode of ELC was used initially to
explore wide ranges of values for the period, epoch of inferior conjunction of
the primary $T_{\rm IC,1}$, inclination, mass ratio, primary velocity 
semiamplitude, and Roche lobe filling factor for each star. The Roche lobe
filling factor is defined by \citet{oro00} as the ratio of the radius of the 
star toward the inner Lagrangian point ($L_{\rm 1}$) to the distance to 
$L_{\rm 1}$ from the center of the star, $f \equiv x_{\rm point}/x_{L \rm 1}$.

The use of ELC included fixing the temperature of each star to the values
found in the tomographic reconstructions (\S \ref{tomog}) and also fixing 
the radius ratio ($R_1 / R_2$) based upon the temperatures, surface fluxes, and
monochromatic flux ratio of the two stars. Non-zero eccentricities for the 
HD~42401 system were explored during fitting but rejected based on the 
higher $\chi^2$ values for those fits. 

To estimate the uncertainties based on our best fit, the values of the seven
fitted parameters were varied in the calculation of $\sim 3~\times~10^6$ 
light and radial velocity curves. The well-explored $\chi^2$ surface was then
projected as a function of each fitted parameter or astrophysical parameter of
interest. The lowest $\chi^2$ value is found for each parameter, and the 
1-$\sigma$ uncertainty may be estimated by the region where 
$\chi^2~\le~\chi^{2}_{\rm min}~+1$. These values and uncertainties are listed
in Table \ref{orb} for the orbital parameters and Table \ref{ELCparms} for 
the astrophysical parameters. Also listed are $R_{\rm eff}$, the effective
radius of a sphere with the same volume, $R_{\rm pole}$, the polar radius
of each star, and $R_{\rm point}$, the radius of the each star toward the inner
Lagrangian point. Our derived period of P= 2.18706 $\pm$ 0.00005 days for 
HD~42401 agrees well with the value found by the \textit{Hipparcos} satellite 
of 2.18709 $\pm$ 0.00050 days \citep{esa97}.


\section{Discussion and Conclusions}

The goal with this work was to determine the masses and radii of the stars
in the HD~42401 system to great accuracy. I have determined the masses of the
two stars to $1.1 \%$ and $0.6 \%$ and the radii of the stars to $0.7 \%$ and 
$0.8 \%$, for the primary and secondary respectively. Armed with these numbers 
and the results from Table \ref{ELCparms}, I can consider the details of the 
HD~42401 system.

\begin{figure}
\begin{center}
{\includegraphics[angle=90,height=12cm]{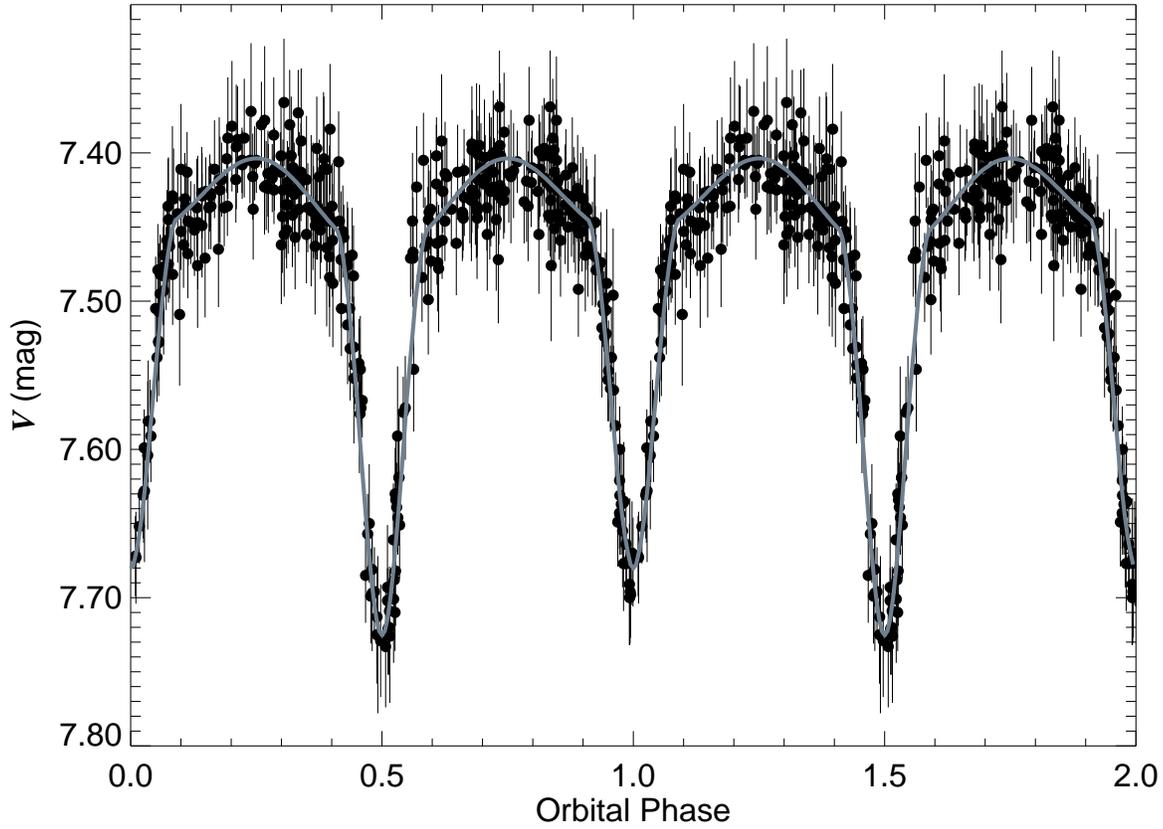}}
\end{center}
\caption{$V$-band light curve for HD~42401. These data were taken from the All 
  Sky Automated Survey database \citep{poj02} and are presented here in phase 
  according to our best fit solution. The model is the thick gray line 
  and the data are the filled dots with $V$ uncertainties represented by line 
  segments. Phase zero corresponds to inferior conjunction of the primary 
  star.}
\label{fig4}
\end{figure}

\begin{figure}
\begin{center}
{\includegraphics[angle=90,height=12cm]{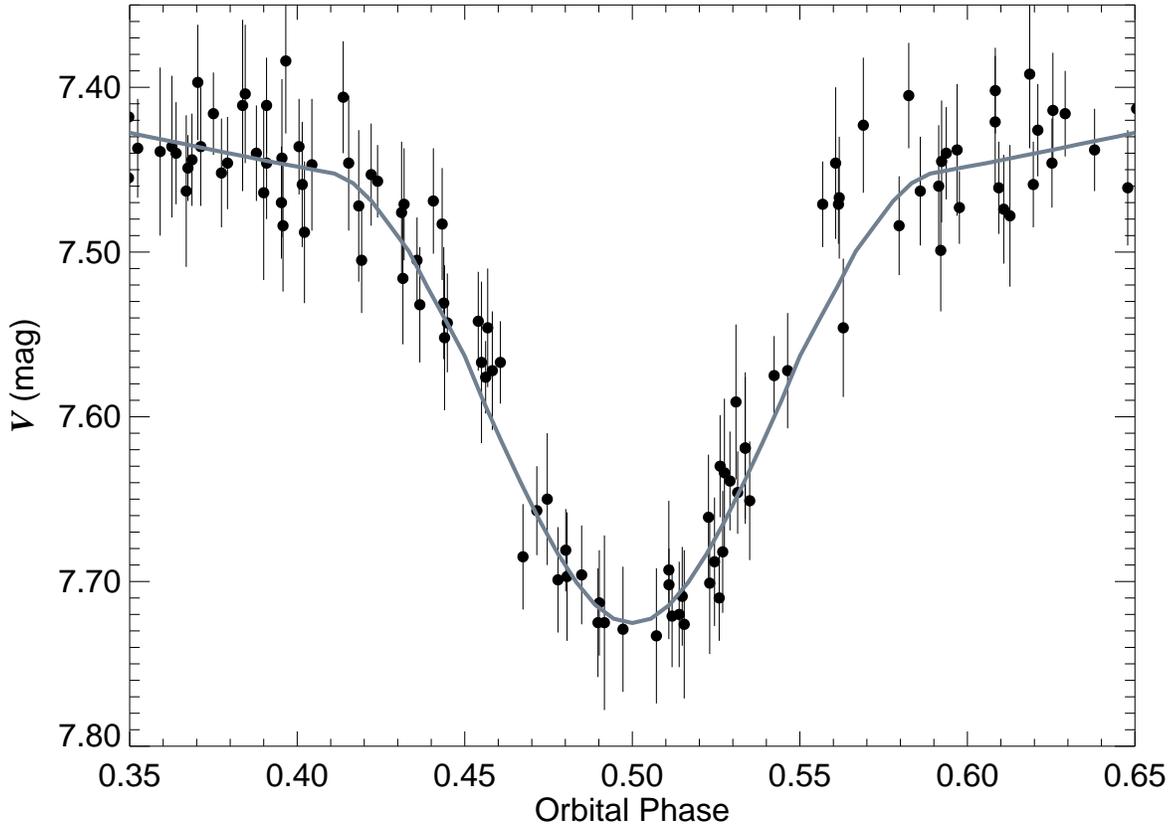}}
\end{center}
\caption{$V$-band light curve for HD~42401 around the time of primary eclipse. 
  As with Fig. \ref{fig4}, the model is the thick gray line and the data are 
  filled dots with line segments representing uncertainties. Residuals are 
  distributed as expected around the best fit.}
\label{fig5}
\end{figure}

\begin{figure}
\begin{center}
{\includegraphics[angle=90,height=12cm]{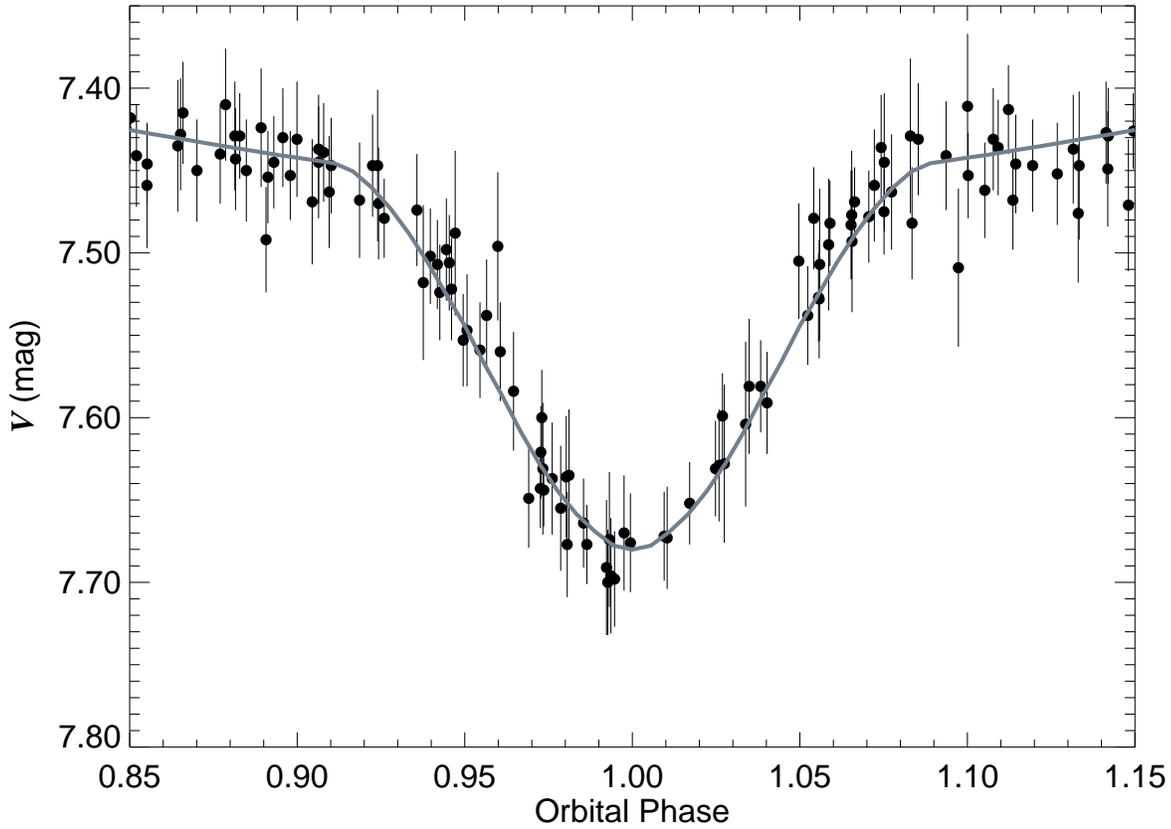}}
\end{center}
\caption{$V$-band light curve for HD~42401 around the time of secondary 
  eclipse. As with Fig. \ref{fig4}, the model is the thick gray line and the 
  data are filled dots with line segments representing uncertainties. Here, as 
  in Fig. \ref{fig5}, the residuals are distributed as expected around the 
  best fit.}
\label{fig6}
\end{figure}

As is seen in Table \ref{ELCparms}, both stars are well within their Roche
radii but experience tidal distortion that is evident in the light curve
(Fig. \ref{fig4}). The rotational velocities derived from the 
tomographic reconstructions (Table \ref{tomo}) match very well with the 
synchronous rotation values found by ELC (Table \ref{ELCparms}), indicating 
that this system has achieved synchronous rotation, and is therefore not very
young. 

\begin{figure}
\begin{center}
{\includegraphics[angle=90,height=12cm]{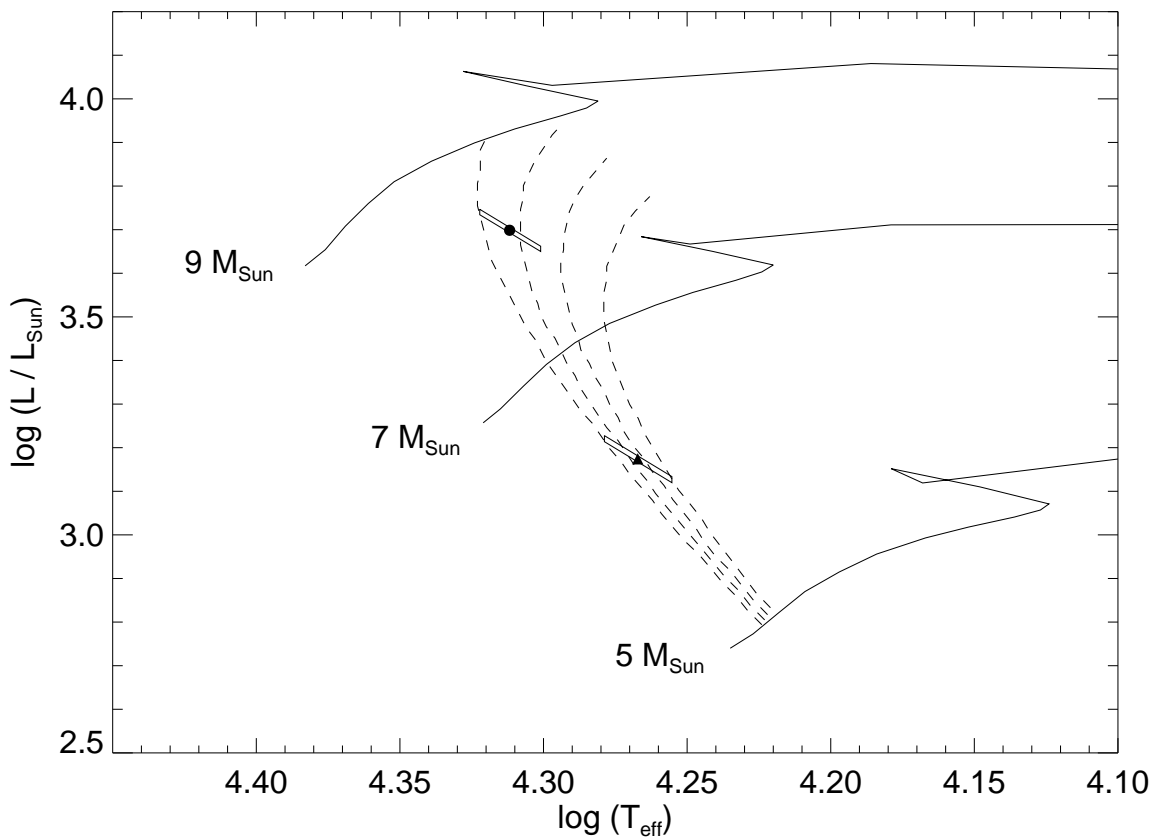}}
\end{center}
\caption{A theoretical H-R diagram showing the location of the primary star 
  (\textit{filled circle}) and secondary star (\textit{filled triangle}) of 
  HD~42401 including uncertainty regions for each. Also plotted are 
  evolutionary tracks for stars of various masses 
  from \citet{sch92} and isochrones (\textit{vertical dashed lines}) from 
  \citet{lej01} for solar metallicity with ages of 21.9, 25.1, 27.5, and 31.6 
  Myr going from left to right. The positions of the two stars are consistent 
  with an age of $\sim$25 Myr.}
\label{fig7}
\end{figure}

To make an estimate of the age of the system, I used the effective 
temperatures from Table \ref{tomo} and radii from Table \ref{ELCparms} to plot
the two stars of HD~42401 on a theoretical H-R diagram and compare their 
locations to
evolutionary tracks. The result is shown in Figure \ref{fig7}, plotted against
evolutionary tracks for stars of 5, 7, and 9 $M_{\odot}$ from \citet{sch92} 
as well as isochrones from \citet{lej01} for solar metallicity with ages of
21.9, 25.1, 27.5, and 31.6 Myr. The location of the stars is most consistent 
with an age of $\sim$25 Myr. The model tracks shown in Figure~\ref{fig7} are 
for non-rotating stellar models. The present ratios of spin angular velocity
to critical angular velocity are $\Omega / \Omega_{\rm crit} = 0.45$ and
0.31 for the primary and secondary, respectively (assuming synchronous 
rotation), and the evolutionary tracks for such moderate rotation rates are
only slightly steeper and more extended in time than those illustrated
\citep{eks08}. Thus, our derived age may slightly underestimate the actual
value. 

Both stars appear (in Fig. \ref{fig7}) to be overluminous for the derived 
masses of 7.42 $M_{\odot}$ for the primary and 5.16 $M_{\odot}$ for the 
secondary. Table 3 of \citet{har88} lists astrophysical parameters for stars
as a function of spectral type for main sequence stars based on empirical data
from eclipsing binaries. The mass and effective 
temperature of the primary fit between the listed values for spectral types
B2 (mean of 8.6 $M_{\odot}$) and B3 (mean of 6.1 $M_{\odot}$), but the radius
is much larger than the means for comparable spectral types and matches a 
B0.5 star (mean of 5.5 $R_{\odot}$). The mass of the secondary 
is consistent with the B4 spectral type (mean of 5.1 $M_{\odot}$) while the
effective temperature and radius appear more consistent with the B3 spectral 
type. In a study of eclipsing 
binaries in the Small Magellanic Cloud, \citet{hil05} found several systems
of comparable mass that, like HD~42401, are overluminous compared to model
predictions. However, our findings for HD~42401 seem to conflict with the 
results of \citet{mal03} who shows that early B-type stars that are in close 
systems and rotate more slowly than single stars are on average smaller than 
those same single stars. In a subsequent paper, \citet{mal07} studied 
well-separated binaries in an effort to use the properties of their component 
stars for a more direct comparison with
single stars. His resulting mass-luminosity-radius relations, when applied to
our results for HD~42401, predict less luminous, hotter and smaller components.
This is perhaps not surprising, due to the age of HD~42401 and the evolution
of its components from the zero age main sequence.

\begin{figure}
\begin{center}
{\includegraphics[angle=90,height=12cm]{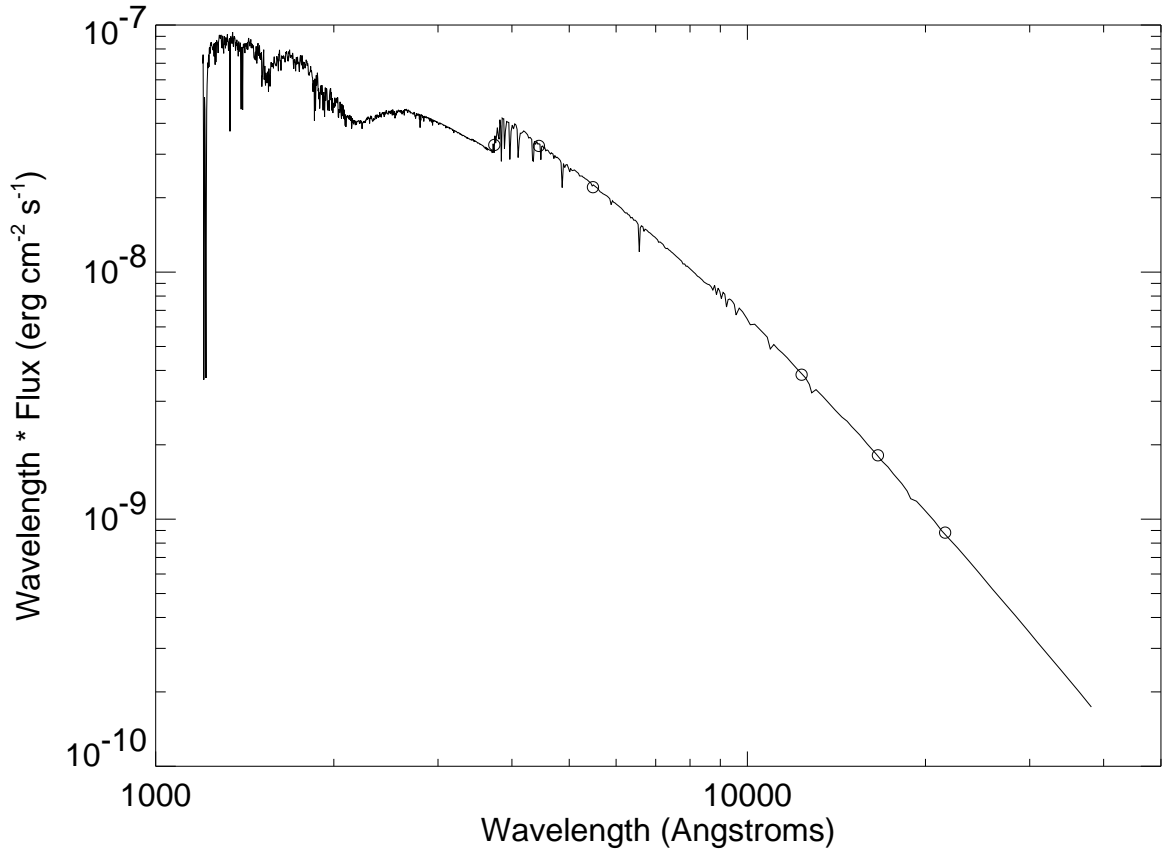}}
\end{center}
\caption{The spectral energy distribution and fit for the combined light of the
  HD~42401 components (\textit{solid line}) to Johnson $U, B, V, J, H, K_s$ 
  photometry (\textit{open circles}).}
\label{fig8}
\end{figure}

I can also estimate the distance to the system by fitting a spectral energy
distribution (SED) to various photometric measurements. The spectra were not
flux calibrated, so to create an SED, I must rely on photometry performed on 
the system.
HD~42401 is a relatively bright $(V \sim 7.4)$ system and has thus been well
studied. The problem arises in estimating the phase at which a particular
observation in the historical literature was made. Two observations of HD~42401
were taken by $IUE$ at $\phi = 0.12 \pm 0.07$ based on our orbital ephemeris. 
The uncertainty in this estimate comes from the uncertainties in our values
for $T_{\rm IC,1}$ and period and the number of orbits between the $IUE$ 
observations and our observations. The range in uncertainty for the orbital
phase of the $IUE$ data
brings it close to eclipse, and this was supported by initial SED fits 
including the $IUE$ data that indicated a lower UV flux than indicated by other
measurements. There are various other measurements that also were 
sufficiently close to eclipses to be omitted in SED fitting. Fortunately, 2MASS
\citep{cut03} measurements were obtained at  $\phi = 0.16 \pm 0.03$ and are 
therefore far enough away from
eclipse, within uncertainties, to be used. The only other points used were 
Johnson $UBV$ magnitudes. To obtain these values, I used the value at 
quadrature from our light curve of $V = 7.40$ and applied colors of 
$(B-V) = -0.042$ and $(U-B) = -0.620$ which are averages of several 
observations that are listed by \citet{mer94}. The SED fit is shown in Figure
\ref{fig8} along with the $U, B, V, J, H, K_s$ magnitudes. I used two model 
spectra from \citet{lan07} matching our values in Table \ref{tomo} for the 
effective temperatures and gravities of each star, and scaled in the blue by 
the flux ratio of $0.25 \pm 0.05$ found from the tomographic reconstructions. 
This fit of the SED results in a limb darkened, angular 
diameter for the primary of $\theta_{\rm LD} = 62.7 \pm 1.1~\mu$as with a
reddening of $E(B-V) = 0.15 \pm 0.01$ mag and a ratio of total-to-selective 
extinction of $R = 3.24 \pm 0.10$. By directly comparing this angular 
diameter with the value for the radius of the primary, I estimate the distance
of the system to be $d = 832 \pm 89$ pc. This distance and reddening are 
in excellent agreement with the values in \citet{bow08} of 
$E(B-V) = 0.15 \pm 0.02$ mag and $d = 0.8$ kpc. 

HD~42401 has Galactic coordinates of $\ell = 197\fdg64$ and $b = -3\fdg33$ 
\citep{ree05}. This is close to the Galactic open cluster NGC~2169 at 
$\ell = 195\fdg61$ and $b = -2\fdg93$ (separation $\sim 2\fdg07$). 
\citet{abt77} found that the
earliest spectral type of the cluster members of NGC~2169 was B2 III. 
\citet{jef07} find a reddening value of $E(B-V) = 0.20 \pm 0.01$ mag and a 
distance of $\sim1060$ pc to NGC~2169. Proper motions for the objects are
similar as well, with HD~42401 having 
$\mu_{\alpha}~cos~\delta = -2.25 \pm 1.41$ mas yr$^{-1}$ and 
$\mu_{\delta} = -2.71 \pm 0.53$ mas yr$^{-1}$ \citep{esa97} and NGC~2169 
having $\mu_{\alpha}~cos~\delta = -2.17 \pm 1.41$ mas yr$^{-1}$ and 
$\mu_{\delta} = -2.52 \pm 0.53$ mas yr$^{-1}$ 
\citep{kha05}. The systemic velocity derived here for HD~42401 is $\sim$ 14 km
s$^{-1}$ and the average of 8 stars in NGC~2169 is 22 km s$^{-1}$ 
\citep{kha05}. 
While the proximity of HD~42401 to NGC~2169 is intriguing and the fact that 
the earliest star has a similar spectral type to the components of HD~42401, 
actual cluster membership is unlikely. The age of NGC~2169 in the literature
ranges from $9 \pm 2$ Myr \citep{jef07} to 7.8 Myr \citep{kha05}, too young to 
comfortably fit with the data for the components of
HD~42401 in Figure \ref{fig7}. \citet{kha05} determined the corona radius of
NGC~2169 by matching stellar densities of the cluster to that of the background
and found an angular size of 0.15 degrees. The distance on the sky of HD~42401
from NGC~2169 is about 14 times this radius. 
The difference in distances to the two objects 
and their separation on the sky make cluster membership even more unlikely. 
However, it is possible that a wave of star formation associated with the Orion
arm of the Galaxy swept through the region, leaving HD~42401 behind as it moved
on and eventually led to the creation of NGC~2169. 


\acknowledgments

I would like to thank the anonymous referee whose comments improved the
paper.
I would also like to thank Daryl Willmarth and the staff of KPNO for their 
assistance in making these observations possible and Jerry Orosz for 
clarifying some of the finer points of ELC's 
calculations. I am also greatly indebted to D.~R.~Gies for his suggestions that
improved the quality of this paper. This material is based on
work supported by the National Science Foundation under grants AST 05-06573
and AST 06-06861. I gratefully acknowledge support from the GSU College
of Arts and Sciences and from the Research Program Enhancement fund of the
Board of Regents of the University System of Georgia, administered through
the GSU Office of the Vice President for Research. This research has made 
use of the SIMBAD database, operated at CDS, Strasbourg, France and  
of the WEBDA database, operated at the Institute for Astronomy of the 
University of Vienna.


\bibliographystyle{apj}

\bibliography{apj-jour,paper}


\newpage

\begin{deluxetable}{cccccccc}
\tabletypesize{\scriptsize}
\tablewidth{0pt}
\tablecaption{HD~42401 Radial Velocity Measurements\label{rvs}}
\tablehead{
\colhead{Date}          &
\colhead{Orbital}       &
\colhead{$V_1$}         &
\colhead{$\sigma_{1}$}  &
\colhead{$(O-C)_1$}     &
\colhead{$V_2$}         &
\colhead{$\sigma_{2}$}  &
\colhead{$(O-C)_2$}     \\
\colhead{(HJD$-$2,400,000)}        &
\colhead{Phase}  &
\colhead{(km s$^{-1}$)} &
\colhead{(km s$^{-1}$)} &
\colhead{(km s$^{-1}$)} &
\colhead{(km s$^{-1}$)} &
\colhead{(km s$^{-1}$)} &
\colhead{(km s$^{-1}$)} }
\startdata
 54474.738 & 0.548 & \phn--36.3   & 5.4 & \phn6.9 & \phn\phn90.6 & \phn3.8 & \phn12.9   \\
 54474.836 & 0.593 & \phn--71.7   & 3.3 & --1.3   & \phn138.1    & 13.8    & \phn\phn4.9\\
 54476.690 & 0.441 & \phn\phn82.7 & 0.4 & \phn2.3 & \phn--45.7   & \phn6.3 & \phn20.4   \\
 54476.735 & 0.461 & \phn\phn65.6 & 3.3 & --1.6   & \phn--21.1   & \phn8.3 & \phn18.1   \\
 54476.775 & 0.480 & \phn\phn53.8 & 1.6 & \phn5.7 & \phn--35.8   & 21.7    & --20.9     \\
 54477.672 & 0.890 & \phn--84.6   & 1.8 & --3.8   & \phn152.6    & \phn6.9 & \phn\phn1.2\\
 54477.720 & 0.912 & \phn--70.4   & 4.0 & --6.5   & \phn129.4    & \phn4.7 & \phn\phn2.4\\
 54477.800 & 0.949 & \phn--41.4   & 3.0 & --8.9   & \phn\phn84.4 & \phn2.0 & --13.8     \\
 54478.660 & 0.342 & \phn139.3    & 0.9 & --3.9   & --160.9      & \phn5.4 & \phn\phn8.2\\
 54478.676 & 0.349 & \phn133.3    & 2.0 & --6.3   & --160.1      & \phn5.6 & \phn\phn3.6\\
 54478.718 & 0.368 & \phn125.0    & 1.9 & --3.3   & --156.2      & \phn5.4 & \phn--9.2  \\
 54478.779 & 0.396 & \phn109.6    & 4.1 & \phn0.3 & --114.4      & 11.0    & \phn\phn4.9\\
 54478.840 & 0.424 & \phn\phn87.6 & 2.0 & --1.6   & \phn--76.1   & 16.9    & \phn10.9   \\
 54479.667 & 0.802 & --128.4      & 0.9 & --1.9   & \phn218.7    & \phn4.6 & \phn\phn0.6\\
 54479.718 & 0.825 & --115.9      & 1.2 & \phn2.3 & \phn214.2    & \phn5.3 & \phn\phn8.3\\
 54479.760 & 0.845 & --109.8      & 0.9 & --0.8   & \phn194.8    & \phn2.0 & \phn\phn2.4\\
 54479.825 & 0.874 & \phn--90.2   & 0.5 & \phn1.3 & \phn174.5    & \phn2.6 & \phn\phn7.5\\
 54480.635 & 0.245 & \phn163.8    & 0.7 & --1.5   & --197.1      & \phn6.8 & \phn\phn6.6\\
 54480.684 & 0.267 & \phn163.2    & 3.8 & --1.6   & --196.5      & \phn5.7 & \phn\phn6.2\\
 54480.755 & 0.300 & \phn155.8    & 2.3 & --3.3   & --194.5      & \phn4.2 & \phn--1.0  \\
 54480.802 & 0.321 & \phn151.8    & 2.0 & --0.5   & --186.8      & \phn4.3 & \phn--3.9  \\
 54481.635 & 0.702 & --127.3      & 2.5 & \phn1.6 & \phn240.8    & \phn6.7 & \phn20.8   \\
 54481.680 & 0.722 & --135.7      & 2.4 & --2.7   & \phn226.5    & \phn6.5 & \phn\phn0.0\\
 54481.728 & 0.744 & --135.8      & 0.6 & --1.1   & \phn222.2    & \phn2.5 & \phn--7.4  \\
 54481.794 & 0.775 & --135.3      & 2.7 & --2.5   & \phn234.0    & \phn4.6 & \phn\phn6.9\\
 54483.646 & 0.622 & \phn--94.4   & 2.2 & --3.3   & \phn171.4    & \phn5.9 & \phn\phn8.2\\
 54483.687 & 0.640 & --108.5      & 1.9 & --5.8   & \phn180.2    & \phn2.5 & \phn--0.1  \\
 54483.757 & 0.672 & --120.3      & 2.0 & --1.5   & \phn200.8    & \phn2.9 & \phn--3.6  \\
 54483.839 & 0.710 & --129.4      & 2.0 & \phn1.3 & \phn223.4    & \phn3.6 & \phn\phn0.5
\enddata

\end{deluxetable}

\begin{deluxetable}{lcc}
\tablewidth{0pc}
\tablecaption{Tomographic Spectral Reconstruction Parameters\label{tomo}}
\tablehead{
  \colhead{Parameter} &
  \colhead{Primary}   &
  \colhead{Secondary}}
\startdata
Spectral Type\dotfill       & B2.5 IV-V\tablenotemark{a} & B3 V\tablenotemark{a}\\
$T_{\rm eff}$ (kK)\dotfill  & 20.5 $\pm$ 0.5 & 18.5 $\pm$ 0.5\\
log $g$ (cgs)\dotfill       & \phn3.75 $\pm$ 0.25 & \phn4.00 $\pm$ 0.25\\
$V~\sin~i$ (km s$^{-1}$)\dotfill & 125 $\pm$ 10 & \phn75 $\pm$ 15\\
$F_{\rm 2} / F_{\rm 1}$ (blue) & \multicolumn{2}{c}{0.25 $\pm$ 0.05}
\enddata
\tablenotetext{a}{These spectral types are estimated from derived values of 
$T_{\rm eff}$ and log $g$.}
\end{deluxetable}

\begin{deluxetable}{lc}
\tablewidth{0pc}
\tablecaption{Circular Orbital Solution for HD~42401\label{orb}}
\tablehead{
  \colhead{Element} &
  \colhead{Value}   }
\startdata
$P$~(days)\dotfill                      & \phn\phn\phn\phn2.18706 $\pm$ 0.00005\\
$T_{\rm IC,1}$ (HJD--2,400,000)\dotfill & 54477.9130 $\pm$ 0.0002\\
$K_{\rm 1}$ (km s$^{-1}$)\dotfill       & \phn\phn151.4 $\pm$ 0.3 \\
$K_{\rm 2}$ (km s$^{-1}$)\dotfill       & \phn\phn217.9 $\pm$ 1.0 \\
$\gamma_{\rm 1}$ (km s$^{-1}$)\dotfill  & \phn\phn\phn15.32 $\pm$ 0.07\\
$\gamma_{\rm 2}$ (km s$^{-1}$)\dotfill  & \phn\phn\phn12.9 $\pm$ 0.2\\
rms~(primary)~(km s$^{-1}$)\dotfill     & \phs\phs3.7 \\
rms~(secondary)~(km s$^{-1}$)\dotfill   & \phs\phs9.7 \\
rms~(photometry)~(mag)\dotfill          & \phs\phs0.007 \\
\enddata
\end{deluxetable}

\begin{deluxetable}{lcc}
\tablewidth{0pc}
\tablecaption{ELC Model Parameters for HD~42401\label{ELCparms}}
\tablehead{
  \colhead{Parameter}  &
  \colhead{Primary}    &
  \colhead{Secondary}  }
\startdata
Inclination (deg)\dotfill & \multicolumn{2}{c}{75.5 $\pm$ 0.2}\\
$M~(M_{\odot})$\dotfill   & \phn\phn7.42 $\pm$ 0.08 & \phn5.16 $\pm$ 0.03\\
$R_{\rm eff}~(R_{\odot})$\dotfill & \phn\phn5.60 $\pm$ 0.04 & \phn3.76 $\pm$ 0.03\\
$R_{\rm pole}$\tablenotemark{a}~($R_{\odot}$)\dotfill & \phn\phn5.40 $\pm$ 0.02 & \phn3.70 $\pm$ 0.01\\
$R_{\rm point}$\tablenotemark{b}~($R_{\odot}$)\dotfill & \phn\phn5.97 $\pm$ 0.02 & \phn3.87 $\pm$ 0.01\\
$V_{\rm sync}~\sin~i$ (km s$^{-1}$)\dotfill & 125.5 $\pm$ 0.9 & 84.3 $\pm$ 0.6\\
log $g$ (cgs)\dotfill & \phn\phn3.812 $\pm$ 0.005 & \phn3.999 $\pm$ 0.006\\
Filling factor\dotfill & \phn\phn0.674 $\pm$ 0.005 & \phn0.588 $\pm$ 0.010\\
$a_{\rm tot}~(R_{\odot})$\dotfill& \multicolumn{2}{c}{16.48 $\pm$ 0.05} \\
$F_{\rm 2} / F_{\rm 1}$ (blue)\dotfill& \multicolumn{2}{c}{\phn0.27 $\pm$ 0.04}
\enddata
\tablenotetext{a}{Polar radius.}
\tablenotetext{b}{Radius toward the inner Lagrangian point.}
\end{deluxetable}


\clearpage

%
%

\input{epsf}

\input{epsf}

\input{epsf}

\input{epsf}

\input{epsf}

\input{epsf}

\input{epsf}

\input{epsf}

\end{document}